\documentclass[12pt]{amsart}
\usepackage{graphicx,amssymb}
\usepackage{amsfonts}

\begin{document}
\title[Spherically Symmetric Solutions in Macroscopic Gravity]{Spherically Symmetric Solutions in\\ Macroscopic Gravity}
\author{R. J. van den Hoogen}
\address{R. J. van den Hoogen, Department of Mathematics, Statistics, and Computer Science,
St. Francis Xavier University, \\ Antigonish, Nova Scotia, Canada, B2G 2W5}
\email{rvandenh@stfx.ca}

\date{\today}

\begin{abstract}
Schwarzschild's solution to the Einstein Field Equations was one of the first and most important solutions that lead to the understanding and important experimental tests of Einstein's theory of General Relativity.  However, Schwarzschild's solution is essentially based on an ideal theory of gravitation, where all inhomogeneities are ignored.  Therefore, any generalization of the Schwarzschild solution should take into account the effects of small perturbations that may be present in the gravitational field.  The theory of Macroscopic Gravity characterizes the effects of the inhomogeneities through a non-perturbative and covariant averaging procedure.  With similar assumptions on the geometry and matter content, a solution to the averaged field equations as dictated by Macroscopic Gravity are derived.   The resulting solution provides a possible explanation for the flattening of galactic rotation curves, illustrating that Dark Matter is not real but may only be the result of averaging inhomogeneities in a spherically symmetric background.
\end{abstract}

\keywords{Averaging in General Relativity, Dark Matter, Macroscopic Gravity}


\maketitle


\section{Introduction}

The most common assumption in the study of modern cosmology and gravity is that the gravitational field is described by Einstein's Theory of General Relativity which is assumed to be valid on all length scales. Cosmologies and isolated gravitational models based on this assumption have been very successful in explaining a wide range of observational phenomenon. For example, on length scales of the solar system, the bending of light curves, the precession of perihelia, the gravitational red-shift, and the radar echo delay all provide significant evidence that Einstein's Theory of General Relativity is adequate to describe the gravitational dynamics in our own solar system \cite{Will}.  Furthermore, the observed expansion of the universe, the observed abundance of light elements created in the first few moments after the Big Bang, and the observed relic Cosmic Microwave Background radiation all attest to the validity of Einstein's theory of General Relativity on much larger scales.

However, everything is not as rosy as it might seem.  Contrary to Einstein's theory, the universe is accelerating \cite{DE}.  This acceleration cannot be explained in General Relativity without the introduction a mysterious substance labeled ``Dark Energy'', a key ingredient in the popular and widely accepted $\Lambda$CDM model of the universe. The existence and properties of Dark Energy are still a matter of debate, but a different solution may be more attractive, that is, perhaps an alternative theory of gravity ought to be used when modeling large scale gravitational systems such as the universe.  The question then becomes, can General Relativity provide an accurate description of the gravitational field on all length scales?

Indeed, even on galactic scales, both Newtonian gravity and Einstein's theory of General Relativity struggle to resolve the observed rotation curves of spiral galaxies\cite{DM}. There is currently no way to explain the flatness of the rotation curves as the distance from the galactic center increases without introducing a substance that has been named ``Dark Matter'', the second ingredient in the popular $\Lambda$CDM model. Dark Matter is a novel and unknown material that does not emit any light, and therefore is difficult to observe directly. However, even the premise of dark matter has its problems; it is sometimes modeled as a cloud of pressureless dust, and therefore, the mass distribution of the dark matter would behave as $r^{-3}$, whereas a solution to the flatness of the rotation curves would really require a mass distribution having a profile of $r^{-2}$. Again, perhaps an alternative and more natural solution exists which explains the dynamics of gravitational systems on intermediate scales, i.e., on the scale of galaxies.

Both the universe and a galaxy can be loosely described as systems of point-like or discrete particles.  For the universe, the constituent particles would be galaxies or galaxy clusters while for a galaxy the constituent particles would be the stellar systems within. Because we are dealing with inhomogeneous distributions of point-like particles in each case, the problem of determining the gravitational dynamics (inside the universe or inside a galaxy) is one that requires either a relativistic N-body simulation or an averaging procedure that averages both sides of the Einstein Field equations on some appropriate scale.

Given that General Relativity has been tested to a high degree of accuracy on scales of the solar system \cite{Will}, it is therefore reasonable to assume that any averaging procedure used to model large scale gravitational effects ought to be based on Einstein's theory of General Relativity. There have been several approaches that purport to average gravitational fields based on Einstein's Theory of General Relativity, (see \cite{Ellis,Zalaletdinov1,Zalaletdinov2,Krasinski,OLD_AVERAGE} and references within for a complete discussion).  However, almost all of these approaches are either non-covariant, or perturbative in nature and therefore their general applicability is questionable.   The theory of Macroscopic Gravity as proposed and developed by R. Zalaletdinov  \cite{Zalaletdinov1,Zalaletdinov2} is covariant and non-perturbative by construction and therefore does not suffer the same weaknesses as the other approaches.  The Zalaletdinov approach employs a covariant space-time averaging procedure for microscopic tensor fields that produces an averaged tensor field.  When this averaging procedure is applied to the field equations of General Relativity, one obtains a generalization of the Einstein Field Equations of General Relativity for the averaged spacetime that includes a new tensor field.  This new tensor field, constructed from a connection correlation tensor $\bf Z$ (which is further dependent upon an affine deformation tensor $\bf A$), satisfies an independent set of differential and algebraic equations \cite{Zalaletdinov1,Zalaletdinov2}.

Unfortunately for a variety of reasons including but not limited to the complexity of the equations describing Macroscopic Gravity, the potential of Macroscopic Gravity has not been fully explored; with the exception of \cite{COVAR_SPAT,CPZ,CP} who have made preliminary attempts to understand the theory.  Here, in this paper we will attempt to illustrate some of the effects of the gravitational stress energy tensor due to the connection correlation field in a class of static spherically symmetric macroscopic spacetimes.  The proposed gravitational models can be used to model galaxies in general, and galactic bulges and halos in particular. Reasonable and physically plausible assumptions on the macroscopic geometry and the connection correlation tensor will be made and the complete set of macroscopic gravity equations will be analyzed.  The resulting solutions will not be solutions to the Einstein Field equations, but will be solutions to the Averaged Field Equations of Macroscopic Gravity.  We will show that the effective energy density due to the connection correlation field has a mass distribution profile of $r^{-2}$, thereby providing an alternative explanation of the flattened galactic rotation curves.

We shall use the same notation as \cite{Zalaletdinov2} for all quantities.  For quick reference, $G_{\alpha\beta}$ is the metric of the averaged or macroscopic spacetime having a Levi-Cevita connection $\overline {\mathcal F}^\alpha{}_{\beta\gamma}\equiv \langle {\mathcal F}^\alpha{}_{\beta\gamma} \rangle$, where ${\mathcal F}^\alpha{}_{\beta\gamma}$ are the bilocal extensions of the Levi-Cevita connection of the microscopic spacetime.  The Riemann curvature tensor corresponding to the macroscopic connection $\overline {\mathcal F}^\alpha{}_{\beta\gamma}$ is denoted as $M^{\alpha}{}_{\beta\gamma\delta}$ with a Ricci tensor $M_{\alpha\beta}=M^{\mu}{}_{\alpha\mu\beta}$.  Covariant differentiation with respect to connection $\overline {\mathcal F}^\alpha{}_{\beta\gamma}$ is denoted with $||$ and any index that is underlined is not in included in the antisymmetrization.  The connection correlation tensor ${\bf Z}$, is defined as the difference between the average of the product
 and the product of the averages of the bilocal extensions of the microscopic connection, via
$$Z^{\alpha}{}_{\beta\mu}{}^{\gamma}{}_{\delta\nu} = \langle {\mathcal F}^{\alpha}{}_{\beta[\mu} {\mathcal F}^{\gamma}{}_{{\underline\delta}\nu]} \rangle
- \overline {\mathcal F}^\alpha{}_{\beta[\mu}\overline {\mathcal F}^\gamma{}_{{\underline\delta}\nu]}.
$$
Please see papers \cite{Zalaletdinov1,Zalaletdinov2} for a complete description of all equations and quantities involved.  Given a coordinate system, and macroscopic metric $G_{\alpha\beta}$, many tensorial objects are greatly simplified by choosing an orthonormal vector bases or tetrad \cite{Plebanski}, $\{{\bf e}_{(\alpha)}\}$ (parentheses surrounding indices are used to denote tetrad indices) adapted to the metric such that the set of scalars representing the metric tensor
$$\eta_{(\alpha)(\beta)}=e_{(\alpha)}^{\mu}e_{(\beta)}^{\nu}G_{\mu\nu}$$
are constant and take on values $\eta_{(\alpha)(\beta)}={\rm diag}[1,1,1,-1].$
Units are chosen so that $c=G=1$ with $\kappa = 8\pi$.


\section{The Macroscopic Gravity Equations}

\subsection{Assumptions} \label{assumptions}

Due to the difficulty in dealing with the extreme size of the tensorial objects involved in Macroscopic Gravity, we are encouraged to make reasonable geometric assumptions about all objects involved in order to make some progress into the understanding of the structure of the Macroscopic Gravity equations. One must recall that only when geometric assumptions were made about the spacetime (i.e., spherically symmetric and static) in General Relativity did one make progress in obtaining solutions of the
Einstein Field Equations and insight into their physical meaning.

\subsubsection*{Assumption 1: The Averaging Procedure and Length Scale} As outlined above we shall assume that the appropriate averaging procedure to use on large scales is the non-perturbative and covariant procedure as developed by Zalaletdinov \cite{Zalaletdinov1}.  In this paper, the averaging length scale is assumed to be significantly larger than the size of our solar system, but much smaller than the size of typical galaxies (for example, on the order of 10--20 parsecs).

\subsubsection*{Assumption 2: The Metric and Higher Order Connection Correlations} The metric correlations as outlined in \cite{Zalaletdinov1} are assumed to be zero.  In addition, we shall assume that the average of the inverse microscopic metric is equal to the inverse of macroscopic metric.  This should be expected and is not considered unusual in any spacetime with a high degree of symmetry.  Furthermore, all higher order connection correlations ({\bf except} $\bf Z$) of Macroscopic Gravity  are assumed to be zero, this can be done in a self-consistent way as outlined in \cite{Zalaletdinov1}.

\subsubsection*{Assumption 3: Macroscopic Geometry}  For the present moment we are interested in investigating the Macroscopic Gravity equations in static spherically symmetric spacetimes exterior and interior to some matter source. The assumption of spherical symmetry and staticity is consistent with the usual assumptions made about the geometry of the dark matter halo surrounding a galaxy. The macroscopic metric, $G_{\alpha\beta}$, can therefore be expressed without loss of generality, as
\begin{equation}
ds^2 = G_{\alpha\beta}dx^\alpha dx^\beta =r^2 d\theta^2+r^2\sin^2\theta d\phi^2+ e^{2\lambda(r)}dr^2   -e^{2\nu(r)} dt^2.
\label{metric}
\end{equation}
The metric, equation (\ref{metric}), is invariant under a four-dimensional group of Killing vectors; the Killing vector, ${\bf X} = \frac{\partial}{\partial t}$, and the three dimensional rotation group.
This spacetime also allows one to define a timelike unit vector field,
$
u^\alpha = [0,0,0,e^{-\nu(r)}]
$, which has a non-zero acceleration, but has zero vorticity, zero shear, and zero expansion.
The tetrad basis corresponding to metric (\ref{metric}) that will be used in future presentations is
\begin{eqnarray}
& & {e}_{(1)}^{\ \mu}=[\frac{1}{r},0,0,0],\quad\ \quad {e}_{(2)}^{\ \mu}=[0,\frac{1}{r\sin\theta},0,0],\label{tetrad}\\
& &{e}_{(3)}^{\ \mu} =[0,0,e^{-\lambda(r)},0],\quad {e}_{(4)}^{\ \mu} =[0,0,0,e^{-\nu(r)}].\nonumber
\end{eqnarray}

\subsubsection*{Assumption 4: Invariance of Macroscopic Gravity Objects}    We shall assume that the connection correlation tensor $\bf Z$ is invariant under the same group of Killing vectors as the macroscopic metric.  This assumption ensures that ${\bf Z}$ is compatible with the geometry of the macroscopic space.  Note, $\bf Z$ need not be invariant under the same group of Killing vectors as the macroscopic spacetime a priori.   In addition, we shall assume that the affine deformation tensor $\bf A$ has tetrad components that are functions of the coordinate $r$ only.

\subsubsection*{Assumption 5: Averaged Microscopic Matter}   We shall also assume that the average of the microscopic stress energy tensor can be modeled as a perfect fluid, that is
\begin{equation}
\langle t^{\alpha {\rm \ (micro)}}_\beta\rangle = \rho^{\rm (mat)}u^\alpha u_\beta+p^{\rm (mat)}(\delta^\alpha_{\beta}+u^\alpha u_\beta)
\end{equation}
where $\rho^{\rm mat}$ and $p^{\rm mat}$ are the energy density and pressure for the averaged matter and where $u^{\alpha}$ can now also be interpreted as the average four-velocity of the fluid.  We shall further assume that the energy density and the pressure of the averaged matter vanishes in the exterior of the mass producing the gravitational field.  While in the interior of the mass generating the gravitational field, the energy density of the averaged matter is assumed to be constant and non-zero, that is,
\begin{equation}
\rho^{\rm (mat)}=\left\{ \begin{array}{ll} \rho_0,\qquad & r\leq R \\
0, & r>R
\end{array}\right.
\end{equation}
where $R$ is the size of the gravitational source.  These assumptions on the averaged microscopic matter will yield the usual interior and exterior Schwarzschild solutions in the absence of connection correlations.

\subsubsection*{Assumption 6: Electric part of the Connection Correlation Tensor}  We shall assume that the electric part of the connection correlation tensor (see \cite{RR2006,Zalaletdinov3}) is zero, that is, $ZE^\alpha{}_{\beta}{}^{\gamma}{}_{\delta\mu}=Z^\alpha{}_{\beta\mu}{}^{\gamma}{}_{\delta\nu}u^{\nu}=0$. The primary motivation for setting $ZE^\alpha{}_{\beta}{}^{\gamma}{}_{\delta\mu}=0$ is that the constraint equation (\ref{ZZ}) is identically satisfied.
\begin{eqnarray}
Z^{\delta }{}_{\beta \lbrack \gamma }{}^{\theta
}{}_{\underline{\kappa }\pi }Z^{\alpha }{}_{\underline{\delta
}\epsilon }{}^{\mu }{}_{\underline{\nu } \sigma ]}+
Z^{\delta
}{}_{\beta \lbrack \gamma }{}^{\mu }{}_{\underline{\nu } \sigma
}Z^{\theta }{}_{\underline{\kappa }\pi }{}^{\alpha }{}_{\underline{
\delta }\epsilon ]}+
Z^{\alpha }{}_{\beta \lbrack \gamma }{}^{\delta }{}_{ \underline{\nu
}\sigma }Z^{\mu }{}_{\underline{\delta }\epsilon }{}^{\theta
}{}_{\underline{\kappa }\pi ]}+\nonumber 
\\
\quad Z^{\alpha }{}_{\beta \lbrack \gamma }{}^{\mu
}{}_{\underline{\delta } \epsilon }Z^{\theta }{}_{\underline{\kappa
}\pi }{}^{\delta }{}_{\underline{ \nu }\sigma ]}+
Z^{\alpha
}{}_{\beta \lbrack \gamma }{}^{\theta }{}_{ \underline{\delta
}\epsilon }Z^{\mu }{}_{\underline{\nu }\sigma }{}^{\delta
}{}_{\underline{\kappa }\pi ]}+
Z^{\alpha }{}_{\beta \lbrack \gamma }{}^{\delta
}{}_{\underline{\kappa }\pi }Z^{\theta }{}_{\underline{\delta }
\epsilon }{}^{\mu }{}_{\underline{\nu }\sigma ]}=0.
\label{ZZ}
\end{eqnarray}

The assumptions {\bf 1 -- 6} greatly simplify the extremely complex system of Macroscopic Gravity Equations, allowing us to analyze the effects of the gravitational stress-energy tensor (due to connection correlations) in the averaged gravitational field equations.

\subsection{The Connection Correlation Tensor}

The connection correlation tensor $\bf Z$, with coordinate components $Z^\alpha{}_{\beta\mu}{}^\gamma{}_{\delta\nu}$, having index symmetries
\begin{equation}
Z^{\alpha}{}_{\beta\mu}{}^{\gamma}{}_{\delta\nu}
=-Z^{\alpha}{}_{\beta\nu}{}^{\gamma}{}_{\delta\mu}
= Z^{\gamma}{}_{\delta\nu}{}^{\alpha}{}_{\beta\mu}
= -Z^{\gamma}{}_{\delta\mu}{}^{\alpha}{}_{\beta\nu}\label{symm}
\end{equation}
can be uniquely represented by the set of scalars ${Z}^{(\rho)}{}_{(\sigma) (\chi) }{}^{(\phi)}{}_{(\kappa)(\psi)}$ through the use of the tetrad (\ref{tetrad}).
Assuming that ${\bf Z}$ is static and invariant under rotations places a significant number of
constraints on the number of independent components.  We find that after all symmetry conditions are imposed, only 150 independent components of $\bf Z$ remain, each of which is a function of the coordinate $r$ only.
With the assumption that $ZE^\alpha{}_{\beta}{}^{\gamma}{}_{\delta\mu}=0$, we find that there are only 75 independent components of $\bf Z$ remaining.
The algebraic cyclic identities,
\begin{equation}
Z^{\alpha}{}_{\beta[\mu}{}^{\gamma}{}_{\delta\nu]}=0
\end{equation}
yield an additional 50 linearly independent constraints.
The differential cyclic constraint equations,
\begin{equation}
Z^{\alpha}{}_{\beta[\mu}{}^{\gamma}{}_{{\underline\delta}\nu||\sigma]}=0
\end{equation}
can be sub-divided into two parts, an algebraic part (containing no derivatives), and a differential part.
The algebraic part of the differential cyclic identities yields an additional 22 algebraic constraints thereby leaving us with only 3 independent components of $\bf Z$ remaining.  The differential part of the differential cyclic identities yields three differential equations for each of the three remaining independent variables. The remaining differential equations are easily solved with the result that the remaining independent tetrad components of $\bf Z$ behave as ${Z}^{(\rho)}{}_{(\sigma) (\chi) }{}^{(\phi)}{}_{(\kappa)(\psi)}(r) \propto r^{-2}$.   Introducing three constants of integration $h_1,h_2,h_3$, the 32 nontrivial components of ${\bf Z}$ are
\begin{eqnarray*}
Z^{(1)}{}_{(1)(1)}{}^{(2)}{}_{(1)(2)} & = & \frac{h_1}{r^2}\\
Z^{(1)}{}_{(1)(1)}{}^{(1)}{}_{(2)(2)} & = & \frac{h_1}{r^2}\\
Z^{(2)}{}_{(2)(2)}{}^{(2)}{}_{(1)(1)} & = & \frac{h_1}{r^2}\\
Z^{(2)}{}_{(2)(2)}{}^{(1)}{}_{(2)(1)} & = & \frac{h_1}{r^2}\\
Z^{(3)}{}_{(1)(1)}{}^{(3)}{}_{(2)(2)} & = & \frac{h_2}{r^2}\\
Z^{(4)}{}_{(1)(2)}{}^{(4)}{}_{(2)(1)} & = & \frac{h_2}{r^2}\\
Z^{(4)}{}_{(1)(2)}{}^{(3)}{}_{(1)(1)} & = & \frac{h_3}{r^2}\\
Z^{(4)}{}_{(2)(2)}{}^{(3)}{}_{(2)(1)} & = & \frac{h_3}{r^2}\\
\end{eqnarray*}
where the remaining 24 nontrivial components are obtained via the symmetries of {\bf Z} (see equation \ref{symm}).

\subsection{The Affine Deformation Tensor}

The equations of Macroscopic Gravity also involve a set of equations that determines an affine deformation tensor $\bf A$ (see \cite{Zalaletdinov1,Zalaletdinov2} for additional details).
The affine deformation tensor satisfies the algebraic equation
\begin{equation}
A^{\epsilon}{}_{\beta[\rho}R^{\alpha}{}_{\underline{\epsilon} \sigma
\lambda]}-A^{\alpha}{}_{\epsilon[\rho}R^{\epsilon}{}_{\underline{\beta}
\sigma \lambda]}=0,\label{AR}
\end{equation}
and the differential equation
\begin{equation}
A^{\alpha}{}_{\beta[\sigma||\rho]}-A^{\alpha}{}_{\epsilon[\rho}A^{\epsilon}{}_{\underline{\beta}\sigma]}=-\frac{1}{2}Q^{\alpha}{}_{\beta\rho\sigma}.
\label{DA}
\end{equation}
where
$$R^\alpha{}_{\beta\gamma\delta}=M^\alpha{}_{\beta\gamma\delta}+Q^\alpha{}_{\beta\gamma\delta}$$
is a non-Riemannian curvature tensor and where
$$
Q^{\alpha}{}_{\delta\mu\nu} = 2 Z^\alpha{}_{\epsilon\mu}{}^{\epsilon}{}_{{\delta}\nu}.
$$
Equation (\ref{AR}) yields 32 constraints and therefore there are only 8 independent components of $\bf A$ remaining.  The algebraic part equation (\ref{DA}) yields an additional 7 algebraic constraints and therefore there is only one independent component of $\bf A$ remaining.  Solving the differential part of equation (\ref{DA}) yields the last remaining component.  The nontrivial tetrad components of $\bf A$ are completely determined to be
\begin{eqnarray*}
A^{(1)}{}_{(1)(1)}&=&\frac{4\sqrt{h_1}}{r}\\
A^{(1)}{}_{(2)(2)}&=&\frac{2\sqrt{h_1}}{r}\\
A^{(2)}{}_{(1)(2)}&=&\frac{2\sqrt{h_1}}{r}\\
A^{(4)}{}_{(4)(4)}&=& {\mathcal A}_0e^{-\nu(r)}
\end{eqnarray*}
where ${\mathcal A}_0$ is a constant of integration.  We observe that the sign of $h_1$ is now restricted,
\begin{equation}
h_1\geq 0.\label{restrict}
\end{equation}

\subsection{The Gravitational Stress-Energy Tensor}

Given our assumptions, the gravitational stress-energy tensor of macroscopic gravity $T_{\beta }^{\alpha {\rm\  (grav)}}$ due to
connection correlations \cite{Zalaletdinov1,Zalaletdinov2},
\begin{equation}
\kappa T_{\beta }^{\alpha {\rm\ (grav)}}= (-Z^{\alpha }{}_{\mu \nu \beta }+\frac{1}{2}\delta _{\beta }^{\alpha
}Q_{\mu \nu })G^{\mu \nu }, \label{T_grav}
\end{equation}
can now be determined, where
$$
Z^{\alpha }{}_{\mu \nu \beta }=2Z^{\alpha}{}_{\mu
\epsilon}{}^{\epsilon }{}_{\nu \beta },\quad{\rm and}\quad Z^{\epsilon }{}_{\mu
\nu \epsilon }=Q_{\mu \nu }.$$
The nontrivial components of $\kappa T_{\beta }^{\alpha {\rm\
(grav)}}$ are
\begin{eqnarray*}
{\kappa T}^{(1) {\rm \ (grav)}}_{(1)} = {\kappa T}^{(2) {\rm \ (grav)}}_{(2)}&=& 0\\
{\kappa T}^{(3) {\rm \ (grav)}}_{(3)}&=& -4\frac{h_{1}}{r^2} \\
{\kappa T}^{(4) {\rm \ (grav)}}_{(4)}&=& -4\frac{h_{1}}{r^2} \\
\end{eqnarray*}
The effective energy density of the gravitational stress-energy tensor of macroscopic gravity is $$\kappa\rho^{(\rm grav)} \equiv \kappa T_{\beta }^{\alpha {\rm\ (grav)}} u_{\alpha}u^{\beta}= \frac{4h_1}{r^2}\geq 0.$$ We note that the sign of $\rho^{(\rm grav)}$ is determined by equation (\ref{restrict}).  The effective energy density exterior and interior to a static spherically symmetric matter source have the same functional form; in both cases the effective energy density behaves as $r^{-2}$. The effective pressure of the gravitational stress-energy tensor of macroscopic gravity has only a radial component, $$\kappa p^{(\rm grav)}_{radial}= -\frac{4h_1}{r^2} = -\kappa\rho^{(\rm grav)}\leq 0$$ which the macroscopic gravity equations determines to be equal in magnitude, but opposite in sign to the effective energy density.  If one interprets these results using a fluid formulation, one finds that the gravitational stress-energy tensor of macroscopic gravity can be effectively modeled as an anisotropic fluid, with non-trivial radial and zero tangential pressures.  Before proceeding further, it is worth mentioning that the stress-energy tensor of macroscopic gravity $T_{\beta }^{\alpha {\rm\  (grav)}}$ satisfies the Weak, Dominant, and Strong energy conditions \cite{HawkingEllis}.

Coley and Pelavas in a series of two papers \cite{CP} have observed something similar; however, the process and perspective they followed and what is used in this paper are different.  They made assumptions about the form of the microscopic geometry and after making a reasonable assumption on the nature of the inhomogeneities, explicitly calculated (using Zalaletdinov's approach to averaging) the spatially averaged value of the Einstein Field Equations to explicitly determine the form of the gravitational correlation tensor.  In particular, they started with the assumption that the microscopic geometry is described by a spherically symmetric metric and determined that the gravitational correlation term behaves as an imperfect fluid.  Here in this paper, a different perspective is used, a macroscopic perspective; no assumption is made about the form of the microscopic geometry, nor any explicit assumptions on the nature of the inhomogeneities.  The only assumption on the geometry is made at the macroscopic level, and the effects of the inhomogeneities have been incorporated directly into the connection correlation tensor.  In this paper, the microscopic geometry may be ``close'' to being spherically symmetric and static, in that it will have a group of ``almost Killing'' vectors, and only after averaging do these ``almost Killing'' vectors become true Killing vectors of the macroscopic geometry.

\subsection{The Averaged Field Equations}

The averaged Einstein Field Equations \cite{Zalaletdinov1,Zalaletdinov2} are
\begin{equation}
E^\alpha{}_{\beta} \equiv G^{\alpha \epsilon }M_{\epsilon \beta }-\frac{1}{2}\delta _{\beta}^{\alpha }G^{\mu \nu }M_{\mu \nu }
=-\kappa \langle {\bf t}_{\beta}^{\alpha {\rm (micro)}}\rangle
-\kappa T_{\beta }^{\alpha {\rm
(grav)}} \label{MG_eqns}
\end{equation}
together with the conservation equation for the averaged microscopic matter
\begin{equation}
\langle {\bf t}_{\beta}^{\alpha {\rm (micro)}}\rangle{}_{||\alpha} = 0.
\end{equation}

Without assuming anything about the form of the averaged energy density, we obtain the following system of differential equations,
\begin{eqnarray}
E^1{}_{1}= E^2{}_{2} &=& e^{-2\lambda}\left(\frac{\lambda'}{r}-\frac{\nu'}{r}-\nu''+\lambda'\nu'-(\nu')^2\right) = -\kappa p^{\rm (mat)},\label{eq1}\\
E^3{}_{3} &=&e^{-2\lambda}\left(\frac{e^{2\lambda}}{r^2}-\frac{1}{r^2}-\frac{2\nu'}{r}\right) = 4\frac{h_{1}}{r^2}-\kappa p^{\rm (mat)},\label{eq2}\\
E^4{}_{4} &=&e^{-2\lambda}\left(\frac{e^{2\lambda}}{r^2}-\frac{1}{r^2}+\frac{2\lambda'}{r}\right) = 4\frac{h_{1}}{r^2}+\kappa\rho^{\rm (mat)} ,\label{eq3}
\end{eqnarray}
and the conservation equation
\begin{equation}
{p^{\rm (mat)}}'=-(\rho^{\rm (mat)}+p^{\rm (mat)})\nu'.\label{eq4}
\end{equation}


\section{Exterior Solution}
For $r>R$ we have $\langle {\bf t}_{\beta}^{\alpha {\rm (micro)}}\rangle=0$, (i.e., $\rho^{\rm (mat)}=p^{\rm (mat)}=0$).
Equation (\ref{eq4}) is identically satisfied and equation (\ref{eq3}) is easily solved to obtain
\begin{equation}
e^{-2\lambda(r)}=(1-4h_1)\left(1-\frac{2\widetilde M}{r}\right)
\end{equation}
where $\widetilde M$ is an arbitrary constant of integration.  We note that for $r>2\widetilde M$ we have the constraint $ h_1\leq 1/4$. Subtracting equation (\ref{eq3}) from equation (\ref{eq2}) we find that
\begin{equation}
\nu'=-\lambda'
\end{equation}
which implies that
\begin{equation}
e^{2\nu(r)}=Ce^{-2\lambda(r)}
\end{equation}
where $C$ is a second arbitrary constant of integration.  However, we are able to set $C$ to any value through a re-scaling of the coordinate $t$.   Here, for simplicity, we shall set $C=(1-4h_1)^{-1}$ and write the macroscopic metric of the exterior of a static spherically symmetric source as
\begin{equation}
ds^2 =r^2\, d\theta^2+r^2\sin^2\theta\, d\phi^2+ \frac{1}{(1-4h_1)}\left(1-\frac{2\widetilde M}{r}\right)^{-1}\,dr^2   -\left(1-\frac{2\widetilde M}{r}\right) dt^2
\label{metric2}
\end{equation}
which has the usual coordinate singularity at $r=2\widetilde M$ and reduces to the standard Schwarzschild solution when $h_1=0$ \footnote{Technically, there is still a two parameter family of connection correlations, but these connection correlations have no effect on the averaged geometry.}.  Recall, for the standard Schwarzschild metric, the Ricci scalar is trivial, however, for the averaged macroscopic geometry determined by the averaged field equations and explicitly written as equation (\ref{metric2}), we find that the Ricci scalar $R=\frac{8h_1}{r^2}$ clearly showing that the geometry determined by equation (\ref{metric2}) is different than Schwarzschild.

It is instructive to rewrite the line element (\ref{metric2}) in isotropic coordinates
\begin{eqnarray*}
x &=& \bar r ^{1/\sqrt{1-4h_1}}\sin\theta\cos\phi,\\
y &=& \bar r ^{1/\sqrt{1-4h_1}}\sin\theta\sin\phi,\\
z &=& \bar r ^{1/\sqrt{1-4h_1}}\cos\theta,\\
&\ & \mbox{\rm where}\\
r &=& \bar r\left(1+\frac{\widetilde M}{2\bar r}\right)^2
\end{eqnarray*}
which yields a line element of the form
\begin{equation}
ds^2 =\left(1+\frac{\widetilde M}{ 2\bar r}\right)^4 \bar r^{2-2/\sqrt{1-4h_1}}\left(dx^2+dy^2+dz^2\right)-    \left(\frac{1-\frac{\widetilde M}{ 2\bar r}}{1+\frac{\widetilde M}{ 2\bar r}}\right)^2 dt^2.
\label{metric3}
\end{equation}
The macroscopic spacetime described by line-elements (\ref{metric2}) or (\ref{metric3}) do not approach the Minkowski spacetime in the limit as $r \to \infty$ or $\bar r \to \infty$ (i.e, far field limit), nor should we expect it to, since the spacetime is not vacuum, but contains non-trivial corrections from averaging.  The usual weak-field interpretation is not appropriate here because the weak-field limit assumes that the gravitational field is described by Newtonian Gravity, which we have assumed is only appropriate on much, much smaller scales.  The General Relativistic limit, and the consequent weak-field interpretation at infinity only takes place when $h_1=0$.


\section{Interior Solution}

We begin by defining the mass function $\widetilde m(r)$ as
\begin{equation}\displaystyle
(1-4h_1)\widetilde m(r) =\frac{\kappa}{2}\int_0^r \rho^{\rm (mat)} s^2\, ds
                 = \left\{ \begin{array}{cc}
                    \frac{\kappa\rho_0r^3}{6} & {\rm for}\quad 0<r<R \\
                    \frac{\kappa\rho_0R^3}{6} & {\rm for}\quad  R\leq r
                  \end{array}\right.
\end{equation}
For $r\geq R$ we observe that the mass function is a constant, $$\widetilde m(r)=\frac{1}{(1-4h_1)}\frac{\kappa\rho_0R^3}{6} = \frac{M_{Sch}}{(1-4h_1)}$$ where $M_{Sch}$ is the usual Schwarzschild mass of the matter source in the absence of connection correlations.

Using this mass function for $r\leq R$, equation (\ref{eq3})  can be integrated to yield
$$
e^{-2\lambda(r)} = (1-4h_1)\left(1 - \frac{2\widetilde m(r)}{r}\right) =
(1-4h_1)\left(1-\frac{r^2}{\widetilde R_0^2}\right)
$$
where $\widetilde R_0^2 = \frac{3(1-4h_1)}{\kappa\rho_0}$.
Equation (\ref{eq4}) yields
\begin{equation}
p^{\rm (mat)} = c_1e^{-\nu(r)} - \rho_0 \label{eq5}
\end{equation}
and where $c_1$ is an arbitrary constant that will be determined by the boundary conditions.  Substituting equation (\ref{eq5}) into equation (\ref{eq2}) one obtains a differential equation for $\nu'$.  Integrating, we find
\begin{equation}
e^{\nu(r)}=\frac{3c_1}{2\rho_0} - c_2\left(1-\frac{r^2}{\widetilde R_0^2}\right)^{1/2}\label{eq6}
\end{equation}
where $c_2$ is another constant of integration.
Substituting equation (\ref{eq6}) into equation (\ref{eq5}) we find an expression for the pressure
\begin{equation}
p^{\rm(mat)}(r) = \rho_0 \frac{-\frac{1}{2}c_1 + c_2\rho_0\left(1-\frac{r^2}{\tilde R_0^2}\right)^{1/2}}{\frac{3}{2}c_1-c_2\rho_0\left(1-\frac{r^2}{\tilde R_0^2}\right)^{1/2}}.
\end{equation}

We shall match the interior solution to the exterior solution by requiring that the $p^{\rm (mat)}(R)=0$, and that the metric functions, $\lambda(r)$ and $\nu(r)$, are continuous at the boundary $r=R$.  This is accomplished by setting
$$c_2=\frac{1}{2} \qquad {\rm and}\qquad c_1=\rho_0\left(1-\frac{R^2}{{\widetilde R_0}^2}\right)^{1/2}.$$
Therefore, the interior solution of the averaged Einstein Field Equations having constant interior energy density $\rho_0$ is
\begin{eqnarray}
ds^2 &=& r^2\, d\theta^2+r^2\sin^2\theta\, d\phi^2+ \frac{1}{(1-4h_1)}\left(1-\frac{r^2}{\widetilde R_0^2}\right)^{-1}\,dr^2 \\  &&\qquad\qquad\qquad -\frac{1}{4}\left(3\left(1-\frac{R^2}{{\widetilde R_0}^2}\right)^{1/2}-\left(1-\frac{r^2}{{\widetilde R_0}^2}\right)^{1/2}\right)^2 dt^2\nonumber
\end{eqnarray}

The continuity condition on the metric functions at the boundary also allows us to interpret the the constant of integration $\widetilde M$ present in the exterior solution.  We obtain the condition
$$\widetilde M = \frac{M_{Sch}}{1-4h_1}= M_{Sch} +\frac{4h_1}{1-4h_1} M_{Sch}$$
which can be now interpreted as the total gravitational mass due to matter together with a small correction due to averaging.


\section{Discussion}

The flattening of rotation curves of spiral galaxies is the primary argument for the existence of Dark Matter.  This Dark Matter is very often modeled as a pressureless dust in a spherical halo surrounding the galaxy.  However, a pressureless dust yields a density profile that behaves as $r^{-3}$ when in reality, all that is required is a density profile of $r^{-2}$.  It is even more curious to note, that the rotational velocity of objects far from the galactic center continues to be constant, even though it is commonly assumed that this galactic halo of dark matter has some finite boundary, which has yet to be observed.   In the context of Macroscopic Gravity, we have found that effective energy density due to connection correlations is, $\rho_{grav}\propto r^{-2}$, and that it has the same form both inside and outside the mass generating the gravitational field.   Consequently, the energy density due to connection correlations can explain the flattening of rotation curves of spiral galaxies, without the need to add Dark Matter.  A back of the envelope calculation using Newtonian Mechanics\footnote{For now this is just a toy model, the assumption that Newtonian mechanics is adequate for the description of the dynamics of galaxies may be questionable, as the equations for geodesics also change because of the averaging.}, using a typical observed rotational velocity of $200 \ km/s$, for large $r$, yields $h_1 \approx 10^{-7}$.

It is curious to note that the form of the gravitational stress-energy tensor due to connection correlations in the simple model presented here takes on the effective form of an anisotropic fluid which can explain the flat rotation curves.  The possibility of modeling dark matter as an anisotropic fluid in spherically symmetric spacetimes has already been investigated and more importantly has been shown to be an extremely viable alternative to the usual dust models of dark matter \cite{SS_aniso}.   What is new here however, is that Macroscopic Gravity now provides the explanation.  Clearly, dark matter in the traditional sense is no longer required; there is no need to use novel and exotic particles to explain the flat rotation curves.  Dark matter is the result of averaging general relativity over an appropriate length scale, and therefore one should perhaps refrain from using the term ``Dark Matter" since it may promote confusion.

There are still many questions that need to be answered in the context of Zalaletdinov's theory of Macroscopic Gravity.  For example, in order to properly compare the theoretical results with astrophysical observations (including the galactic rotation curves mentioned here), one must properly study the geodesic equations for both massive and massless test particles.  A straightforward calculation shows that the geodesic equation for the exterior metric given by equation (\ref{metric2}) becomes
\begin{eqnarray}
\frac{d^2u}{d\phi^2} + u &=& \underbrace{\frac{kM_{Sch}}{L^2}}\qquad + \qquad \underbrace{3M_{Sch}u^2}\qquad +\qquad \underbrace{4h_1u}\label{geo1}\\
&\  & \mbox{\tiny Newton} \qquad\qquad\qquad  \mbox{\tiny GR \ correction} \qquad\qquad  \mbox{\tiny MG\  correction}\nonumber
\end{eqnarray}
where $u=1/r$, $M_{Sch}$ is the Schwarzschild mass, $L$ is the angular momentum, $k=0$ for massless particles, and $k=1$ for massive particles. The form of the geodesic equation is nearly the same as in the traditional version of the geodesic equation for the Schwarzschild metric.   We note the appearance of an additional term linear in $u$ that is not present in the geodesic equation for the standard Schwarzschild spacetime.  Through a re-scaling of the independent variable ($\tilde \phi = \sqrt{1-4h_1}\phi$), one can show that the geodesic equation (\ref{geo1}) becomes
$$\frac{d^2u}{d\tilde \phi^2} + u =\frac{k{\widetilde M}}{L^2} + 3{\widetilde M}u^2 $$
Since $\widetilde M \approx M_{Sch}$, we find that there would only be a small change in the astrophysical observations.  However, any change to the geodesic equations ought to be thoroughly analyzed.

The analysis of gravitational and cosmological models based on Zalaletdinov's Macroscopic gravity is by no means complete.  Even in the spherically symmetric and static case studied here, the effects of the electric part of the connection correlation tensor $\bf Z$ is yet unknown.  The electric part of the connection correlation tensor may significantly alter or on the other hand may fine tune the solution to the galactic rotation curve problem.  The electric part of the connection correlation tensor may also provide a natural explanation for the acceleration of the universe in a cosmological context.  Furthermore, the assumption that the connection correlation tensor $\bf Z$ is invariant under the same Lie group of Killing vectors as the macroscopic metric, although ensuring compatibility between them, is restrictive.  It will be of great interest to relax this invariance condition and study the effects of more general connection correlations even in the static spherically symmetric case studied here.

Nevertheless, the result presented in this paper illustrates that Zalaletdinov's Macroscopic Gravity provides an extremely promising explanation for the flattening of rotation curves and therefore this averaged theory of gravity merits further investigation.


\section{Acknowledgements}
The author would like to thank Roustam Zalaletdinov for his tremendous patience, understanding, and guidance.  The author would also like to thank Alan Coley for his comments.  {\sc MG Eqs Code}, \cite{MG_EQS_2007}, {\sc GRTensorII}\cite{GRTENSOR} and the computer algebra package {\sc Maple} were used extensively for the calculations present in this paper.  This work was supported through a grant from the Natural Sciences and Engineering Research Council (NSERC) of Canada.


\end{document}